# AN EFFICIENT ANT BASED QOS AWARE INTELLIGENT TEMPORALLY ORDERED ROUTING ALGORITHM FOR MANETS

Debajit Sensarma [1] and Koushik Majumder [1]

[1] Department of Computer Science & Engineering, West Bengal University of Technology, Kolkata, INDIA
debajit.sensarma2008@gmail.com,
koushik@ieee.org

## ABSTRACT

*A Mobile Ad hoc network (MANET) is a self configurable network connected by wireless links. This type of network is only suitable for temporary communication links as it is infrastructure-less and there is no centralised control. Providing QoS aware routing is a challenging task in this type of network due to dynamic topology and limited resources. The main purpose of QoS aware routing is to find a feasible path from source to destination which will satisfy two or more end to end QoS constrains. Therefore, the task of designing an efficient routing algorithm which will satisfy all the quality of service requirements and be robust and adaptive is considered as a highly challenging problem. In this work we have designed a new efficient and energy aware multipath routing algorithm based on ACO framework, inspired by the behaviours of biological ants. Basically by considering QoS constraints and artificial ants we have designed an intelligent version of classical Temporally Ordered Routing Algorithm (TORA) which will increase network lifetime and decrease packet loss and average end to end delay that makes this algorithm suitable for real time and multimedia applications.*

## 1. INTRODUCTION

A Mobile Ad Hoc Network (MANET) is a collection of mobile nodes without having any existing infrastructure and communicates via wireless links. In recent days, large number of MANET routing algorithms has been proposed due to the dynamic nature of the network topology and no centralized control that increases the overhead in route discovery and maintaining the stable route. Besides this, with the growth of internet, the demand for real time and quality of services (QoS) of network has been increased. So, demand of QoS-aware routing is also increased. The major objectives of QoS-aware routing are:-i) finding path from source to destination satisfying user's requirement. ii) optimizing network resource usage and iii) repairing or re-computing the path quickly in case of path break or link failure or unwanted things like congestion, without degrading the level of QoS. Again, centralized algorithms have scalability problems, static algorithms have trouble keeping up-to-date with network changes, and other distributed and dynamic algorithms have oscillations and stability problems. Ant based routing provides a promising alternative to these approaches. Various Ant Colony routing protocols has been defined in the next section [1-5]. Ant colony provides a number of advantages [6] due to the use of mobile agents and stigmergy (a form of indirect communication used by ants in nature to coordinate their problem-solving activities). It provides scalability, i.e. according to the network size the population of the agents can be adapted. It is Fault tolerant, i.e. it does not rely on a centralized control mechanism. Therefore, node mobility or link breakage does not result in catastrophic failure. Besides this, it provides adaptation, where according to the network changes agents can change, die or reproduced. Speed is achieved, because network topology change can be propagated very fast. Agent acts individually, independent of other network layers, so it gives modularity. Also, autonomy is provided because little or no human supervision is required. It also provides parallelism, i.e. agents operate in parallel. However, besides these advantages, one of the biggest difficulties with ant colony algorithms applied in network routing area is that multiple constraints often make the routing problem intractable [7].

The proposed algorithm is a QoS aware multipath on-demand routing algorithm. The advantage of this proposed scheme is that, it supports energy efficient multipath routing as well as taken care of the QoS constraints-delay, bandwidth, energy and drain rate which is very essential for multimedia applications.

The paper is organized as follows: Section 2 illustrates the basic idea behind ant colony optimization. Section 3 of this paper gives the brief information about some related works with this research work. Section 4 describes Temporally Ordered Routing Algorithm (TORA). In section 5 mathematical models are described and in section 6 the proposed algorithm combining the idea of ACO and TORA is illustrated. Section 7 explains the performance analysis. Concluding remarks and future works are given in Section 8.

## 2. ANT COLONY OPTIMIZATION

The ant colony optimization (ACO) meta-heuristic is a generic problem representation and it is based on the behaviour of ants. It adopts real ant's foraging behaviour. Ants initially start random walk when multiple paths exist between nests to food. They lay a chemical substance called pheromone during their food searching trip as well as their return trip to the nest. Pheromone serves as route mark, which the ants follow. Newer ants will take that path which has higher pheromone concentration and also the pheromone concentration of that path will increase by the time. This is an autocatalytic effect and this helps the solution to be emerging quickly [1].

Some properties characterizes ACO instances for routing problems, they are

a. In a network where the topology changes dynamically, highly adaptive routing is necessary. Also, in the network without any centralized control, due to node mobility the link can be broken any time and the communication may be lost. If multiple paths exist between source and the destination, one path lost cannot effect the communication, because anyone of the existing paths can be used for routing. ACO provides both the traffic-adaptive and the multipath routing.
b. It is necessary to choose a path for routing which satisfies both the required constraints for routing and for this, some previous information are needed and based on the newer and the previous information the path is chosen. In ACO, both the passive and active information are gathered and monitored.
c. ACO uses the stochastic components for routing.
d. ACO does not allow local search estimates to have global impact for the required solution. In ACO no routing information has to transmit to neighbour or all the nodes.
e. ACO does not set paths like other greedy shortest path schemes, at the time of path set up it also taken care of load balancing. So, it taken care of the link quality also.
f. Another important aspect is parameter setting. It is done by ACO in less sensitive way.

Figure 1 illustrates the behavior of ants. A set of ants moves along a straight line from their nest S to a food source D (Figure 1a). At a given moment, an obstacle is put across this way so that side (A) is longer than side (B) (Figure 1b). Now, the ants have to decide which direction they will take: either A or B. The first ones will choose a random direction and will deposit pheromone along their way. The ants taking the way SBD (or DBS), will arrive at the end of the obstacle (depositing more pheromone on their way) before those that take the way SAD (or DAS). So, pheromone intensity of route SBD becomes greater than that of route SAD. So, the ants choose the path SBD (Figure 1c). The ants will then find the shortest way between their nest and the food source.

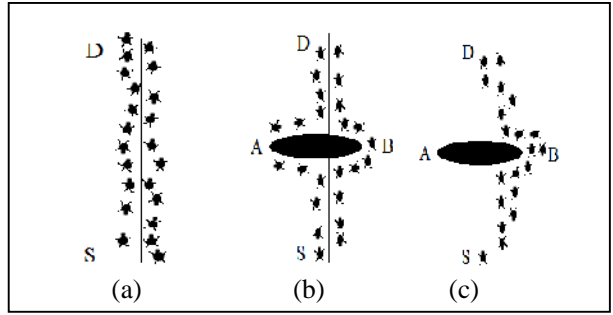

**Fig. 1.** Behavior of ants for searching the food from S to D

In most cases, an artificial ant will deposit a quantity of pheromone represented by $\tau_{i,j}$ only after completing their route and not in an incremental way during their advancement. This quantity of pheromone is a function of the found route quality.

Pheromone is a volatile substance. An ant changes the amount of pheromone on the path $(i, j)$ when moving from node $i$ to node $j$ as follows:

$$\tau_{i,j} = \rho \cdot \tau_{i,j} + \Delta\tau_{i,j} \qquad (1)$$

Where $0 < \rho < 1$ and $\rho$ is the pheromone evaporation factor which avoids infinite increment of pheromone which may leads to stagnation of the route.

At one point i, an ant chooses the point j (i.e. to follow the path $(i, j)$) according to the following probability:

$$P_{i,j} = \frac{(\tau_{ij})^\alpha \cdot (\eta_{ij})^\beta}{\sum_{i,k \in C}(\tau_{ik})^\alpha \cdot (\eta_{ik})^\beta} \qquad (2)$$

Where,
$\tau_{i,j}$: is the pheromone intensity on path $(i, j)$.
$\eta_{i,j}$: is the ant's visibility field on path $(i, j)$ (an ant assumes that there is food at the end of this path).
$\alpha$ and $\beta$: are the parameters which control the relative importance of the pheromone intensity compared to ant's visibility field.
C: represents the set of possible paths starting from point i $((i, k)$ is a path of C).

Like real pheromone the artificial pheromone decreases over time for fast convergence of pheromone on the edges. This happen in ACO according to the following formula:

$$\tau_{i,j} = (1-q) \cdot \tau_{i,j} \qquad q \in (0,1] \qquad (3)$$

International Journal of Computer Networks & Communications (IJCNC) Vol.5, No.4, July 2013

## 3. RELATED WORKS

An Ant Colony-based Multi Objective Quality of Service Routing for Mobile Ad hoc Networks [5] proposed by P. Deepalakshmi et al. is an ant-based multi-objective on-demand QoS routing algorithm for mobile ad hoc network. Here, Link failure is detected quickly as node uses updated view of the network with positive and negative feedback. As it is a multi-path routing algorithm, therefore, it supports node mobility in a better way. The main drawback of this approach is, at the time of route discovery, it only considers delay, throughput and jitter as the QoS metrics but other network layer or MAC layer metrics are not considered for achieving more stable route and high throughput. Furthermore, congestion issues like flow control in the network are not considered and packets may be lost due to the absence of proper flow control mechanism. Also, admission control and resource utilization are not taken into account. Lastly, the MAC layer collision is not considered, which causes the degradation of efficiency of the network.

S. Kanan et al. proposed Ant Colony Optimization for Routing in Mobile Ad-Hoc Networks [4] which is a multi agent ant based routing algorithm for MANET. This is a new technique increases the connectivity of the nodes during high node mobility and for this the QoS constraints like delay is decreased and packet delivery ratio (PDR) is increased. It mainly focuses on the node mobility by only supporting QoS requirements like delay, PDR etc. which is not sufficient for multimedia application. Other network and link layer metrics have to be considered. There are various other issues. This algorithm uses both proactive and reactive routing. Paths are monitored always, so the overhead is increased in routing. It does not take into account the heterogeneous MANET behavior, where the data rate of each link can be different, so flow control is necessary, which is not noticed here. Consequently, there is a lack of Congestion control. Moreover, in MAC layer as nodes communicate through the shared medium; hence, there is a chance of collision, contention and interference. These problems are not taken into account in this algorithm. Furthermore admission control is not tackled properly for better throughput and efficiency.

Ant Colony Based QoS Routing Algorithm for Mobile Ad Hoc Networks [1] is an on-demand QoS routing algorithm proposed by P.Deepalakshmi et al. This algorithm is highly adaptive in nature and mainly reduces the end to end delay in high mobility cases. This is a good routing scheme when node mobility is high. Here in route discovery the minimum QoS requirements in terms of bandwidth, delay and hop count are considered. But the other QoS constraints i.e. other network layer or link layer metrics like energy, jitter, link stability etc. are not considered here. Furthermore, here link failure is not handled properly. One disadvantage of TCP is that, it treats the link failure as congestion in the network. This phenomenon is not taken into account here. MANETs are heterogeneous in nature, so data rates are different for each link. Thus, when the source node sends Route_Request_Ant to all its neighbors, it does not take the data rate into account. Hence, flow control between the nodes is not considered here which might cause congestion in the network. Nodes in MANET communicate with each other through a shared medium. Thus, packet collision can occur in MANET. As a result, packet loss rate is increased, which degrades the throughput. Moreover, interference of channel is not considered separately and admission control is not taken into account which is useful for proper resource utilization.
S.B.Wankhade et al. proposed an on-demand routing algorithm, Route Failure Management Technique for Ant Based Routing in MANET [2]. This algorithm is inspired by the ant colony routing algorithm. Route failure management is the main key of this algorithm. Authors have shown that it has good maintenance scheme and it supports good packet delivery ratio (PDR) with less packet drop and delay. In this technique, at the time of route discovery, only delay, bandwidth, hop count, PDR are taken into account but other network layer metrics or link layer metrics like throughput, jitter, node buffer size etc. are not taken into consideration. Here, in route updation phase, multiple routes are found and in maintenance phase, fuzzy logic is used to

192

find Link Stability Coefficient. So, a small overhead is involved with route updation in some time interval. Furthermore, here flow control is not handled explicitly, so when source generates HANT it does not takes into account the data rate of the link in heterogeneous MANET. Packet collision and channel interference also are not considered. The computation of fuzzy logic in the node may cause premature death to the nodes due to the limited energy of the nodes. For real time and multimedia communication only improvement of packet delivery ratio is not sufficient, because end to end delay, processing delay at each node and also other QoS constraints affects the throughput.

Ant Based Dynamic Source Routing Protocol to Support Multiple Quality of Service (QoS) Metrics in Mobile Ad Hoc Networks [3] proposed by R. Asokan et al. and it performs well in route discovery phase with dynamically changing topology and produces better throughput with low delay variance. This algorithm performs well in case of dynamically changing topology. But it considers only some of the network layer metrics like delay, jitter, energy and throughput. But, only these constraints are not sufficient for real time applications as other network layer or MAC layer metrics are also necessary. In this routing protocol the packet header size is increased with increasing route length due to its source routing nature. Consequently, the routing overhead is also increased. Again flooding of route request may potentially reach all nodes in the network, so bandwidth wastage increases and efficiency degrades. Besides this, it is a collision and contention prone routing protocol. Thus, packet delivery ratio decreases, congestion increases and throughput also become very poor in case of multimedia communication.

Some of the above ant based routing algorithms are hybrid and some of them are on-demand. Our proposed algorithm is ant based and it is an on-demand routing algorithm which supports multipath routing and takes the advantages of both ACO and TORA. This algorithm unlike the above algorithms is based on link reversal concept and it uses destination sequenced routing. Here, nodes only need to know the information about their one hop neighbors. The main advantage is, control messages are localized to very few numbers of nodes where link breakage or topology changes occur. Here link failure is handled very efficiently. Another case, when network partition occur this algorithm is able to detect the partition and erase all invalid routes. The above discussed existing algorithms have not focused on this issue. It also considers the node drain rate as a QoS constraint, which is a very important QoS constraint for avoiding congestion in the network. Moreover, this algorithm selects only those nodes for routing which satisfy the energy constraint.

A detailed literature survey on ant based QoS aware routing and their comparative analysis in MANET can be found in our previous work [8].

## 4. TEMPORALLY ORDERED ROUTING ALGORITHM (TORA)

Temporally Ordered Routing Algorithm (TORA) is first proposed by Park and Corson [9]. It is a highly adaptive, distributed, source initiated on-demand routing protocol based on a link reversal algorithm. It finds multiple loop free paths from source to destination. The main feature of TORA is that the control messages are localized to a very small set of nodes near the occurrence of the topological change and to achieve that, nodes maintain routing information about adjacent nodes. The protocol performs three main functions: route creation, route maintenance, and route erasure. These three functions are facilitated by the use of three different control packets: query (QRY), update (UPD), and clear (CLR) packets. When a node requires a route to a destination, route creation is initiated where QRY and UPD packets are used to establish directions on previously undirected links. This establishment of links results in a destination-oriented directed acyclic graph (DAG) where any packet sent will reach the destination as it is the only node with no

International Journal of Computer Networks & Communications (IJCNC) Vol.5, No.4, July 2013

downstream links. Topological changes caused by node mobility may cause a node to lose all of its downstream links and hence its route to the destination. With the use of UPD packets, route maintenance ensures that the DAG is re-established to become destination-oriented within a finite time. In the event that a network partition is detected, route erasure uses CLR packets to ensure that all invalid routes are erased by setting all links in the partition to be undirected. Each node i∈ N has a height represented by as a quintuple: $H_i = (\tau_i, oid_i, r_i, \delta_i, i)$, where i) $\tau_i$: A time tag indicating the time of link failure represents the reference level. ii) $oid_i$: Originator-id, id of node that defined the reference level. iii) $r_i$: 1 bit used to divide each reference level into 2 sub-levels. iv) $\delta_i$: Integer used to order nodes with respect to a unique reference level and v) i: Unique identifier of the node. The height of the node is defined by two parameters: a reference level and a delta with respect to the reference level. The first three values in the quintuple represent the reference level while the last two values represent the delta. Each node i (other than the destination) maintains a link-state array with an entry $LS_{i,j}$ for each link (i, j) ∈ L, where j is a neighbor of i. The height of node i, $H_i$ and that of its neighbor j, $HN_{i,j}$ determines the direction of the links, and is directed from the higher node to the lower node. If node i has a non-NULL height, it labels the link $LS_{i,j}$ as i) Upstream (UP), if a neighbor j has a height higher than that of node i. ii) Downstream (DN), if a neighbor j has a height lower than that of node i. iii) Undirected (UN), if a neighbor j has a NULL height. If node i has a NULL height and its neighbor j has a non-NULL height, the link $LS_{i,j}$ is labeled as downstream (DN) as its neighbor is considered lower. Initially, all links are undirected thus all nodes in the network have a height set to NULL, $H_i = (\_, \_, \_, \_, i)$. An exception is the destination id (did) whose height is always ZERO, $H_{did} = (0, 0, 0, 0, did)$. Each neighbor of the destination j ∈ $N_{did}$ sets the height $HN_{j,did}$ to ZERO, $HN_{i,did} = (0, 0, 0, 0, i)$ and labels their link $LS_{j,did}$ as downstream. When a node i discovers a new neighbor j ∈ Ni, it establishes the new link (i, j) ∈ L by adding a new height $HN_{i,j}$ and link-state $LS_{i,j}$ entry for neighbor j. If the new neighbor is not the destination, the height entry is set to NULL, $HN_{i,j} = (\_, \_, \_, \_, i)$ and the link-state is labeled as undirected. If the new neighbor is the destination, the height entry and link-state is set accordingly.

## 5. MATHEMATICAL MODEL

For mathematical analysis MANET is represented by a connected undirected graph. Let G (V, E) represents the mobile ad hoc network. Here V denotes the set of network nodes and E denotes the set of bidirectional links. QoS metrics with respect to each link e ∈E is delay (e) and bandwidth (e). With respected to node n ∈V , it is delay (n), energy (n) and drain rate (n) which is the energy dissipation rate of node 'n'. Another QoS metric considered here is hop count. It is important because multiple hops are used for data transmission in MANET. So, it is necessary to find paths with minimum hops. The main motivation of this proposed algorithm is to find path from source to destination which will satisfy the QoS requirements such as delay, bandwidth, energy, drain rate and hop count.

Let, path (i, j) or R is entire path from node i to j where QoS constraints have to satisfied.

From an arbitrary node i to an arbitrary node j, delay, bandwidth, energy, drain rate and hop count is calculated as-

delay (path (i, j )) or D (R) = $\sum_{e \in P(i,j)} delay(e) + \sum_{n \in P(i,j)} delay(n)$

where, delay (path (i,j )) is the transmission and propagation delay of the path(i,j) and delay (n) is the processing and queuing delay of node 'n' on path(i, j).

bandwidth(path(i,j)) or B(R)= $\min_{e \in P(i,j)}$ {bandwidth(e)}

where, bandwidth (e) is the available bandwidth of that link on path(i, j).

energy (path (i, j)) or E (R) = $\min_{n \in P(i,j)}$ { energy (n)}

where, energy (n) is the residual energy of node 'n' on path(i, j).

drain rate (path (i, j)) or DR (R) = $\max_{n \in P(i,j)}$ { drain rate(n)}

where, drain rate (n) is the rate of energy dissipation of node 'n' on path(i, j).
hop count (path (i, j)) or HC (R) = Number of nodes in the path.

### 5.1 CALCULATION OF PHEROMONE

Ant deposits pheromone during traversal of the link for finding a route. The quantity of pheromone it deposited on each link (i, j) along the route R is noted by $\Delta \tau_{i,j}$ and it is a function of global quality of route R. It is expressed by the following equation-

$$\Delta \tau_{i,j} = \frac{B(R)^{\lambda_B} + E(R)^{\lambda_E}}{D(R)^{\lambda_D} + HC(R)^{\lambda_{HC}} + DR(R)^{\lambda_{DR}}} \qquad (4)$$

Here $\lambda_B$, $\lambda_E$, $\lambda_D$, $\lambda_{HC}$ and $\lambda_{DR}$ are the weight factors which indicate the relative significance of the QoS parameters during pheromone update on path (i, j). The quantity of the deposited pheromone is defined only after finding the route.

The pheromone quantity of the link (i, j) is updated according to the equation (1).

### 5.2 CALCULATION OF PATH PREFERENCE PROBABILITY

Path Preference Probability is calculated in each intermediate node as well as source node upon receiving of QRY Reply_Ant.

Suppose current node i receives QRY Reply_Ant from node j for destination d, then the Path Preference Probability is calculated as-

$$P_{ijd} = \frac{[\tau_{ij}]^{\alpha_1}.[D_{ijd}]^{\alpha_2}.[\eta_{ijd}]^{\alpha_3}.[B_{ijd}]^{\alpha_4}.[E_{ijd}]^{\alpha_5}[DR_{ijd}]^{\alpha_6}}{\sum_{k \in N_i}[\tau_{ik}]^{\alpha_1}.[D_{ikd}]^{\alpha_2}.[\eta_{ikd}]^{\alpha_3}.[B_{ikd}]^{\alpha_4}.[E_{ikd}]^{\alpha_5}[DR_{ijd}]^{\alpha_6}} \qquad (5)$$

Here $\alpha_1$, $\alpha_2$, $\alpha_3$, $\alpha_4$, $\alpha_5$ and $\alpha_6$ are the tunable parameters which control the relative weights of pheromone trails, hop count, bandwidth, energy and drain rate respectively.

$N_i$ is the set of neighbors of i and k is the neighbor node of i through which a path to destination is known.

The relative metrics are calculated from source i to destination d via j as-

$D_{ijd} = \dfrac{1}{\text{delay(path (i, d))}}$

$\eta_{ijd} = \dfrac{1}{\text{hopcount(path(i, d))}}$

$B_{ijd}$ = bandwidth (path(i,d))

$E_{ijd}$ = energy (path (i,d))

$DR_{ijd} = \dfrac{1}{\text{drain rate(path(i, d))}}$

Now, source as well as neighbors has multiple paths from source to destination. The path with higher Path Preference Probability is selected for the data transmission.

## 6. PROPOSED ALGORITHM

The proposed algorithm is a multipath on-demand routing algorithm. Here QoS constraints-delay, bandwidth, energy and drain rate are considered which are very essential for multimedia applications.

This algorithm has three phases namely route discovery phase, route maintenance phase and route erasure phase. In route discovery phase, multiple paths which satisfy the required QoS constraints are created and stored in the cache. The route with higher path preference probability is selected for the routing. In route maintenance phase, when a node fails to transmit packets, it will check for the alternate route to the desired destination with better path preference probability. If such path exists, it is chosen for the routing. In the next case, if the failure node has no outbound link to other nodes in the network, then it will send an error message to the source and immediately start local link maintenance. All routes going through the failure node are deleted from the cache. In the next step, when source finds the error message, it will check for other alternate route in the cache with next better path preference probability. If it exits, that route gets selected for routing. Otherwise, a new route discovery phase is started. If network partition is detected during route maintenance phase, then route erasure phase is invoked and all invalid routes are deleted from cache.

Each node k$\in$ N has a height represented by as a quintuple: H$_i$= ($\tau_k$, oid$_k$, r$_k$, $\delta_k$, k ), where i) $\tau_k$: A time tag indicating the time of link failure represents the reference level. ii) oid$_k$: Originator-id, id of node that defined the reference level. iii) r$_k$: 1 bit used to divide each reference level into 2 sub-levels. iv) $\delta_k$: Integer used to order nodes with respect to a unique reference level and v) k: Unique identifier of the node. Each node k (other than the destination) maintains a link-state array with an entry LS$_{k,j}$ for each link (k, j) $\in$ L, where j is a neighbor of k, L is total number of links. The height of node k, H$_k$ and that of its neighbor j, HN$_{k,j}$ determines the direction of the links, and is directed from the higher node to the lower node. RR$_k$ is route-required flag of node k.

### 6.1. ROUTE DISCOVERY PHASE

**HELLO_Ant packet:** HELLO_Ant is broadcasted in every second by every node in the network. It collects available energy, drain rate of neighbor nodes. Based on the size of HELLO_Ant packet and starting, receiving time, current node will calculate the available bandwidth of the outgoing links.

**QRY Request_Ant packet:** It consists of request starting time, available bandwidth, source id, destination id, stack of visited node addresses.

**QRY Reply_Ant packet:** It consists of hop count, delay, energy, drain rate, bandwidth, source id, destination id, stack of node addresses to be visited.

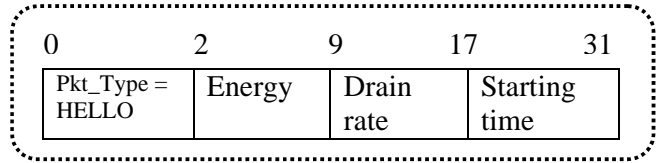

**Fig. 2.** HELLO_ Ant packet format

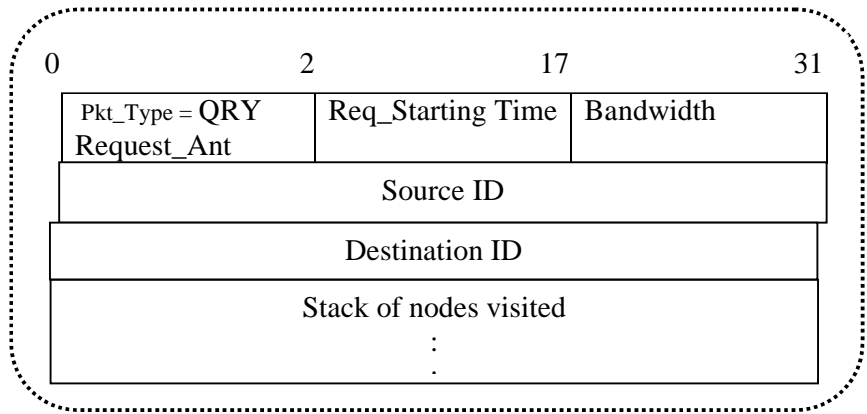

**Fig. 3.** QRY Request_Ant Packet format

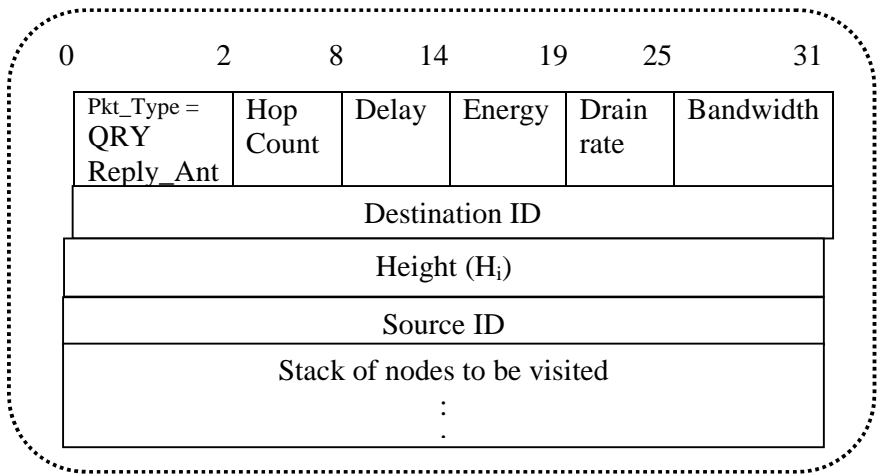

**Fig. 4.** QRY Reply_Ant Packet format

*Algorithm 1: Route Discovery*

**BEGIN**

**Step 1:** Let, Source(S) wants to communicate with the destination (D) with the QoS constraints bandwidth, delay, drain rate, energy. No path exists in the cache previously.

**Step 2:** S sets its height as NULL initially i.e (-,-,-,-, S).

**Step 3:** S sends a QRY Request_Ant packet to its neighbors and if the destination is not one hop away from the intermediate node then node also sets its height as NULL and broadcasts the packet.

**Step 4:** The QRY Request_Ant packet collects the information about bandwidth, delay, energy and drain rate during traversal.

**Step 5:** When an intermediate node k receives a QRY Request_Ant, it has four options:

5.1. If node k has no downstream links and $RR_k$ is unset, it rebroadcasts the QRY Request_Ant packet and sets $RR_k$.

5.2. If node k has no downstream links and $RR_k$ is set, it discards the QRY Request_Ant packet.

5.3. If node k has at least one downstream link and its height is NULL, it sets its height to $H_k = \min \{H_j \mid j \in N_k\} + \{0, 0, 0, 1, 0\}$ and broadcasts a QRY Reply_Ant packet.

5.4. If node k has at least one downstream link and its height is non-NULL, and if a QRY Reply_Ant packet has been broadcast since the link over which the QRY Request_Ant packet was received became active, it discards the QRY Request_Ant packet. Otherwise it broadcasts a QRY Reply_Ant packet. Also, if $RR_k$ is set when a link becomes active, it broadcasts a QRY Request_Ant packet.

**Step 6:** QRY Reply_Ant follows the route corresponding to the QRY Request_Ant but in reverse order.

**Step 7:** When intermediate or source node k receives an QRY Reply_Ant from a neighbor j, k updates $HN_{k,j}$ to reflect the height of node j received in the QRY Reply_Ant and performs one activity from the two options:

7.1. If $RR_k$ is set (implying the height of node k is NULL), node k sets $H_k = \min \{H_j \mid j \in N_k\} + \{0, 0, 0, 1, 0\}$, updates the link $LS_k$, unsets $RR_k$ and broadcasts a QRY Reply_Ant packet with the new information.

7.2. If $RR_k$ is unset, node k updates the links in $LS_k$.
Intermediate nodes also update the pheromone table according to equation (1) and calculate the path preference probability according to equation (5) and also update the probability table, and set the lower level node as downstream node.

**Step 8:** Source also calculates the Path Preference probability.

**Step 9:** If the calculated path preference probability is better than the requirements then the path is accepted and stored in the cache.

**Step 10:** The path with the better Path Preference Probability is selected for data transmission.

**END**

## 6.2. ROUTE MAINTENANCE PHASE

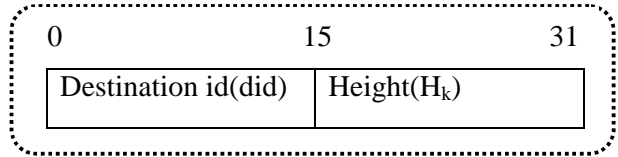

**Fig. 5.** UPD packet format for a node (k)

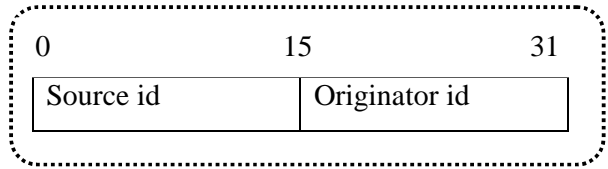

**Fig. 6.** Error packet format for a node

**UPD packet:** It contains the destination id (did), height of the node (say, $H_k$ of node k) that is broadcasting the packet.
**Error packet:** It contains the Source id, Originator id of the node.

A node k is said to have no downstream links if $H_k < HN_{k,j}$ for all non-NULL neighbors $j \in N_k$.

*Algorithm 2: Route Maintenance*

**BEGIN**
Here 2 cases can occur:

**Case 1:** When an intermediate node k detects the route failure, if it has an outbound link ($H_k > HN_{k,j}$) i.e. another unexpired route exists in the cache, the route with better Path Preference Probability is selected for routing.

**Case 2:** When an intermediate node detects the route failure, if it has no outbound link the following sequence of steps are performed:

**Step 1:** The node generates an Error packet and sends to the source node and all intermediate nodes delete routes from their cache which contains that node.

**Step 2:** As soon as source receives the error packet, it deletes routes from the cache which contain the node and if an unexpired route exists in the cache, the route with next better Path Preference Probability is selected for routing.

**Step 3:** After sending the Error packet, the node where link failure occur, starts local link maintenance according to the following conditions:

a. **Generate:** Node k has no downstream links due to link failure.

($\tau_k$, $oid_k$, $r_k$) = (t, k, 0), where t is the time of failure.

($\delta_k$, k) = (0, k).

Node k defines a new reference level, if k has no upstream neighbors, it sets $H_k$ = **NULL**.

b. **Propagate:** Node k has no downstream links due to link reversal following receipt of an UPD and the ordered sets ($\tau_k$, $oid_k$, $r_k$) are not equal for all j ∈ $N_k$.

($\tau_k$, $oid_k$, $r_k$) = max {($\tau_k$, $oid_k$, $r_k$) | j∈ $N_k$}

($\delta_k$, k) = {min { $\delta_j$ | j∈ $N_k$ with ($\tau_j$, $oid_j$, $r_j$) = max {($\tau_j$, $oid_j$, $r_j$) } - 1, k}.

Node k propagates the reference level of its highest neighbors and chooses a reference lower than all neighbors of that reference level.

c. **Reflect:** Node k have no downstream links due to link reversal following receipt of an UPD packet and the sets ($\tau_j$, $oid_j$, $r_j$) are equal with $r_j$=0 for all j∈ $N_k$.

($\tau_k$, $oid_k$, $r_k$) = ($\tau_j$, $oid_j$, 1)

($\delta_k$, k) = (0, k).

Node k reflects back the reference level by setting the r bit.

d. **Detect:** Node i has no downstream links due to link reversal following receipt of an UPD packet and the sets ($\tau_j$, $oid_j$, $r_j$) are equal with $r_j$=1 for all j∈ $N_k$ and $oid_k$=k.

($\tau_k$, $oid_k$, $r_k$) = (_, _, _)

($\delta_k$, k) = (_, k).

Node k has detected a partition and route erasure phase invoked (described below).

e. **Generate:** Node k has no downstream links due to link reversal following receipt of an UPD packet and the sets ($\tau_j$, $oid_j$, $r_j$) are equal with $r_j$=1 for all j ∈ $N_k$ and $oid_j$∈ k.

($\tau_k$, $oid_k$, $r_k$) = (t, k, 0), where t is the time of failure.

($\delta_k$, k) = (0, k).

Node k experienced a link failure between the time it propagated a reference level and the reflected higher sub-level returned from all neighbors. This link failure required no reaction.

**Step 4:** Finally, after performing step 2, step 3, if the cache contains no unexpired routes, again a new route discovery phase is started.

**END**

## 6.3 ROUTE ERASURE PHASE

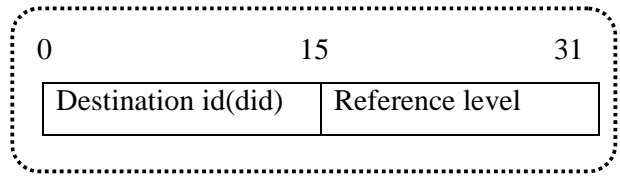

**Fig. 7.** CLR packet format of a node

**CLR packet:** It consists of destination id (did) and the reference level of node. It is used to erase invalid routes.

Route erasure is initiated upon the detection of a partition. Node k sets its height and the height entry for each j ∈$N_k$ to NULL. However, if the destination is a neighbor, the corresponding height array is set to ZERO. Node k then updates its link-state array LS, and broadcasts a CLR packet that consists of a destination id, did and the reflected sub-level of node k, ($\tau_k$, $oid_k$, 1). When a node k receives a CLR packet from a neighbor j∈$N_k$, it reacts according to the following steps of Algorithm 3:

*Algorithm 3: Route Erasure*

**BEGIN**

**Step 1:** If the reference level in the CLR packet matches that of node k, it sets its height and the height entry for each neighbor j∈$N_k$ to NULL (unless the destination is a neighbor, in which case the corresponding height array is set to ZERO), updates all the entries in its link-state array LS, and broadcasts a CLR packet.

**Step 2:** If the reference level in the CLR packet does not match that of node k, it sets the height entry for each neighbor j ∈$N_k$ (with the same reference level in the CLR packet) to NULL and updates the corresponding link-state array entries.

**Step 3:** Lastly, the hop in error is removed from the node's route cache and all routes containing the hop are eliminated.

**END**

## 7. PERFORMANCE ANALYSIS

The proposed algorithm is an adaptive routing algorithm. So, it is suitable for the network where node mobility is higher and no centralized control exists. It takes the advantages of both TORA and Ant Colony Optimization technique.

i) This algorithm is a distributed routing algorithm and each router works autonomously and in ACO no routing table is transmitted to neighbor or all other nodes. So, this decreases the routing overhead.
ii) This algorithm finds multiple paths between source and the destination. Again, ACO always not find shortest path, rather at the time of path set up it also taken care of the QoS requirements and for this, link quality is improved. Besides this, if multiple paths exits and as, ant store paths which satisfies the QoS requirements, losing any one path cannot effect the communication, which is very necessary for the real time and multimedia communication. Use of multi path routing also increases the packet delivery ratio, decreases the packet loss rate. It also utilizes the bandwidth properly and for this, throughput and network stability or lifetime increase.
iii) This proposed algorithm's design is aimed at minimizing the aggregate bandwidth by minimizing the control packets. It also minimizes the communication overhead by localizing algorithmic reaction to topological changes. Especially in case of link failure, the number of nodes that must participate in the reaction is minimized in comparison with the other routing algorithms. In the first phase of route maintenance of the algorithm, if link failure occurs and that time a downstream link exists then only the route with valid QoS requirements are selected for routing. So, it minimizes the number of control packets and available bandwidth utilized properly.
iv) This algorithm performs better in case of dense network. Because, in that case more number of paths are generated and stored in then cache, which also increases reliability by increasing packet delivery ratio.

v) Here, QoS constraints energy of individual nodes and the drain rate of each node are taken into account. Hence, nodes which satisfy the minimum energy requirements participate in the routing. It decreases the packet loss and obviously network becomes more stable.

vi) It is a loop free routing, which reduces the end to end delay and packet loss. Packet delivery ratio automatically increases and so as the throughput.

## 8. CONCLUSION & FUTURE WORKS

QoS aware routing in MANET is a challenging task. Many research works have been carried out in this area. In this paper, an ant based QoS aware multipath routing algorithm is proposed which supports real time and multimedia applications. This algorithm is more adaptive and energy efficient which takes node's remaining energy as well as drain rate (i.e. energy dissipation rate) as QoS parameter. It selects the node which has sufficient resource to satisfy the QoS constraints. In high mobility cases it is very efficient in terms of quick route maintenance. This algorithm takes care of end to end delay, available bandwidth, and hop count as QoS parameter which increases network throughput. It is a multipath routing which increases network stability which is very important for multimedia application as here uninterrupted data transmission is required for increasing the throughput. This routing decreases number of control packets which reduces routing overhead and utilizes bandwidth properly which is also very efficient especially in case of dense networks.

In our future work we will simulate this proposed routing scheme and will compare it with the other QoS aware routing algorithms.

### ACKNOWLEDGEMENTS

The authors would like to thank West Bengal University Technology, Kolkata, India for the supports and facilities provided to carry out this research. The authors also thank the reviewers for their constructive and helpful comments.

**Authors**

Debajit Sensarma has received his B.Sc degree in Computer Science in the year 2009 from university of Calcutta, Kolkata, India and M.Sc degree in computer Science in the year 2011 from West Bengal State University, Kolkata, India. He obtained his M.Tech. degree in Computer Science and engineering from West Bengal University of Technology, Kolkata, India in the year 2013. He is now pursuing PhD from the department of Computer Science and Engineering, University of Calcutta, Kolkata. He has published several papers in International journals and conferences.

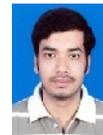

Koushik Majumder has received his B.Tech and M.Tech degrees in Computer Science and Engineering and Information Technology in the year 2003 and 2005 respectively from University of Calcutta, Kolkata, India. He obtained his PhD degree in the field of Mobile Ad Hoc Networking in 2012 from Jadavpur University, Kolkata, India. Before coming to the teaching profession he has worked in reputed international software organizations like Tata Consultancy Services and Cognizant Technology Solutions. He is presently working as an Assistant Professor in the Dept. of Computer Science & Engineering in West Bengal University of Technology, Kolkata, India He has published several papers in International and National level journals and conferences. He is a Senior Member, IEEE.

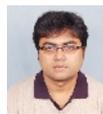